\documentclass[twocolumn]{aastex62}

\usepackage{amsmath}

\usepackage{natbib}

\newcommand\kms{\textrm{ km s$^{-1}$}}
\newcommand\masyr{\textrm{ mas yr$^{-1}$}}

\begin{document}
\title{Runaway Young Stars near the Orion Nebula}
\author[0000-0003-2401-0097]{Aidan McBride}
\affil{Department of Physics and Astronomy, Western Washington University, 516 High St, Bellingham, WA 98225}
\author[0000-0002-5365-1267]{Marina Kounkel}
\affil{Department of Physics and Astronomy, Western Washington University, 516 High St, Bellingham, WA 98225}

\begin{abstract}

The star forming region of the Orion Nebula (ONC) is ideal to study the stellar dynamics of young stars in a clustered environment. Using \textit{Gaia} DR2 we search for the pre-main sequence stars with unusually high proper motions that may be representative of a dynamical ejection from unstable young triple systems or other close three-body encounters. We identify twenty-six candidate stars that are likely to have had such an encounter in the last 1 Myr. Nine of these stars could be traced back to the densest central-most region of the ONC, the Trapezium, while five others have likely interactions with other OB-type stars in the cluster. Seven stars originate from other nearby populations within the Orion Complex that coincidentally scattered towards the ONC. A definitive point of origin cannot be identified for the remaining sources. These observations shed light on the frequency of the ejection events in young clusters.
\end{abstract}

\section{Introduction}

Star clusters are very dynamical regions that contain significantly larger density of stars in a small volume compared to the more diffuse field. As such, stars are much more likely to encounter one another in dense clusters, which could lead to unstable 3-body interactions. This is especially true of the young clusters, as they have not yet settled into a stable configuration, both due to intracluster dynamics, but also in recently formed unstable multiple systems \citep{reipurth2010,schoettler2019}. Such interactions should result in a excitation of velocity of an ejected star, although most are expected to be weak, and would not become unbound from their cluster. Only the strongest interactions would produce runaway stars.

The early proper motion studies in the Orion date back throughout the 20th century \citep[e.g.,][]{strand1958,mcnamara1976,jones1988} relied on the astrometry from the photometric plates that have been observed over the course of up to 80 years. However, as the ONC is located near the galactic anticenter, its proper motions are very small, \added{and without distance measurements for the individual sources it was difficult to distinguish true members from the field stars.} And even with the long temporal baseline, at the resolution of those surveys, it was difficult to precisely measure proper motion of the individual stars\deleted{, and a lack of reliable distance measurements made it difficult to separate the true cluster members from the field stars}. Later, with the launch of Hipparcos, it became possible to track the absolute \replaced{position}{astrometry} of the stars to measure proper motions and parallax simultaneously. But at the distance of the ONC, it was only able to make measurements only for a handful of OB stars in the region, with the relatively poor precision. Despite this, several massive young runaways have been identified in the past. Some notable examples include the pair $\mu$ Col and AE Aur which have been ejected from the Orion Nebula Cluster \citep[ONC;][]{blaauw1961}, and HD 30112 and 43112 which have been ejected from the $\lambda$ Ori cluster \citep{hoogerwerf2001}.

In addition to the proper motions derived from optical observations, significant effort has been made to obtain astrometry in radio regime. The source BN was first identified as runaway through radio interferometry by \citet{plambeck1995}, may have been ejected from the Trapezium \citep{tan2004}, or have been part of a decaying multiple system \citep{rodriguez2005}, has triggered an explosive outflow from passing in close proximity to source I, and this resulted in runaways from the site of the explosion \citep[e.g.][]{luhman2017}.

Other proper motions survey in radio have also been conducted. \citep{dzib2017} collated the observations of the inner ONC over the baseline of almost 30 years with VLA, reaching precision in proper motions of a few \masyr. Additionally, \citep{kounkel2017} have performed a survey of the Orion Complex with VLBA. Due to low sensitivity, only 26 non-thermally emitting stars were detected with a $\sim$0.1 \masyr\ precision, and 3 of these stars (V1961 Ori, V1321 Ori, and Brun 334) were identified as runaways. Similar degree of precision was recently achieved in proper motions derived from the optical and near-infrared from space and ground-based adaptive optics data towards the ONC \citep{kim2019}.

\textit{Gaia} is a successor to Hipparcos, significantly improving on both the precision and sensitivity of the measurements \added{of parallax and proper motion measurements}. Its second data release \citep{gaia-collaboration2018} contains 1.3 \replaced{million}{billion} stars with information on their parallax and proper motions up to $G$ of $\sim$21 mag. It has allowed to constrain the characteristic dynamical state of number of star-forming regions, including the Orion \citep[e.g.,]{kounkel2018a}, and can be used to search for runaways among young stars \citep[e.g., KPNO 15 and 2MASS J04355209+2255039 in Taurus][]{luhman2018}.

In this paper we search for stars that have been ejected through a 3-body interaction using \textit{Gaia} DR2 in the ONC, which is the most massive nearby young cluster and the origin of most previously known ejection events.

\section{Selection of candidate ejected stars}

\begin{figure}
\includegraphics[width=\linewidth]{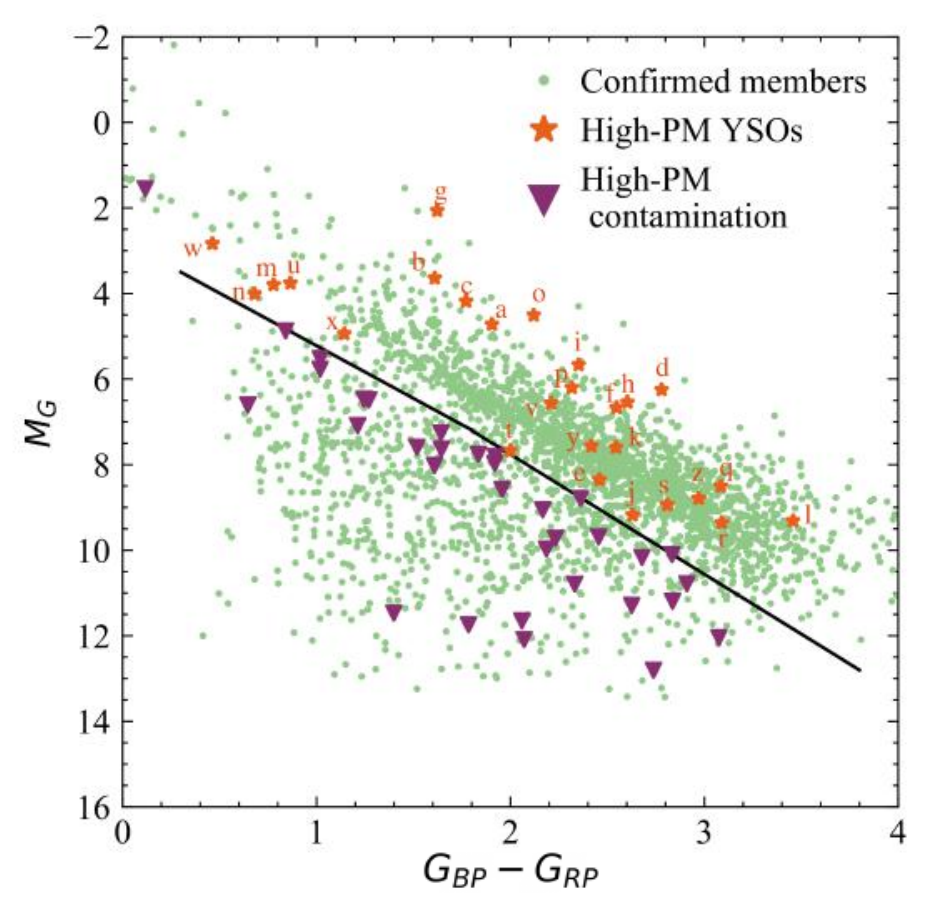}
\includegraphics[width = \linewidth]{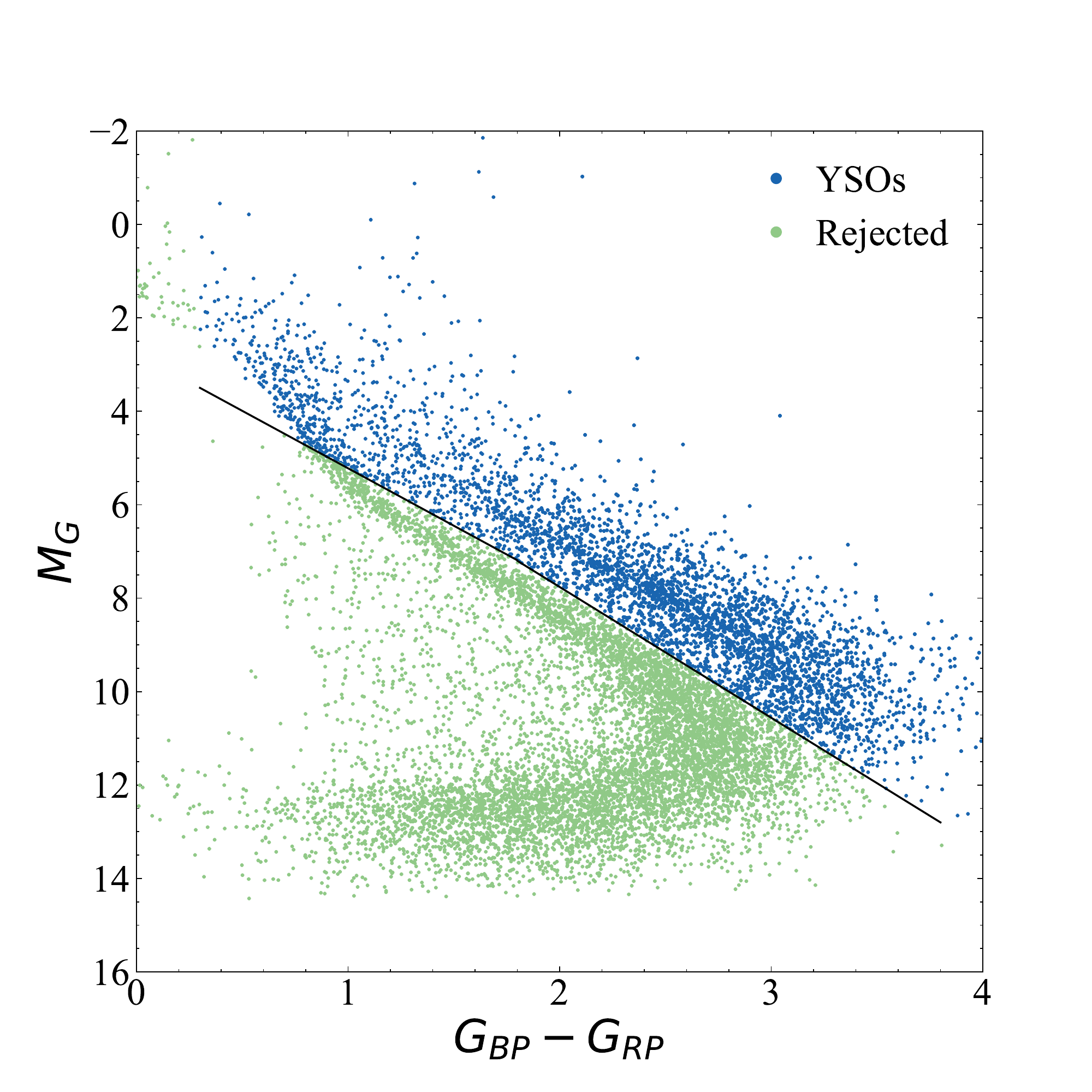}
\caption{Top: HR diagram of confirmed Orion A YSOs. High proper motion sources are shown in red. Some known cluster members have bad photometry or poor parallaxes, and thus appear to be underluminous. $\sim$1--2\% of the sources may be contamination from the field main sequence stars that lie below the photometric cut \added{(39 stars below the photometric cut in a sample of 2995 sources that have \textit{Gaia} astronetry and meet spatial cuts)}. The labels identify the high proper motion sources in the description throughout Section \ref{sec:pms}. Bottom: HR diagram with the photometric cuts used to distinguish between the query-selected YSOs and the field main sequence stars. \label{fig:HR}}
\end{figure}

Ejected stars can be identified by having discrepant proper motions from the cluster mean, however it is necessary to distinguish all of the candidates from the older field stars that have larger velocity dispersion compared to what is found in a young cluster. To do that, we collated a list of confirmed young stars towards the Orion A \added{molecular cloud (of which the ONC is a part of)} from the literature (Table \ref{tab:sources}, \added{references are in the caption}), including those that were identified from near-IR excess, from X-ray emission, confirmed spectroscopically, or by other means. This catalog consists of 5988 stars, of which 3078 are Class I or Class II, and 2910 are Class III. Not all of these sources are detectable with \textit{Gaia}; some may also have poor astrometry and/or photometry due to surrounding nebulosity or nearby companions leading to inaccurately measured parallaxes. The various criteria of youth combined are reliable on the 3$\sigma$ level, with 1--2\% contamination from the main sequence stars (Figure \ref{fig:HR}\deleted{a}). \added{This catalog is not necessarily complete with every single member that has formed within the Orion A, due to the limited field of view of the individual surveys, different sensitivity limits, and various selection effects. However, this catalog should include most of the young stellar objects that have been previously identified towards this region}.

To further clean the sample from the contamination, we also required an independent photometric confirmation of youth for these stars. We downloaded \textit{Gaia} DR2 data towards the ONC, we used the initial cuts to encompass the cluster, centered at (J2000) $\alpha = 83.833^\circ$, $\delta = -5.391^\circ$, with a search radius of 2$^\circ$ and parallax $2 < \pi < 5$, which fully encompasses the cluster. We further restricted this sample using the photometric cuts 
\[M_G<2.46 \times |G_{BP}-G_{RP}| +2.76; 0.3 < |G_{BP}-G_{RP}| <1.8\]
\[M_G<2.8 \times |G_{BP}-G_{RP}| + 2.16; 1.8 < |G_{BP}-G_{RP}|\]
\noindent from \citet{kounkel2018a} to remove the low mass main sequence stars (Figure \ref{fig:HR}\deleted{a}), which selected 4025 stars (Figure \ref{fig:HR}b). This selection is sufficient to remove the low mass stars older than 15--20 Myr. For the final sample we require the intersection between the photometric selection and the membership list, limiting the sample to 1867 stars.

\begin{deluxetable*}{cccccccccc}
\tabletypesize{\scriptsize}
\tablecaption{Known members of the Orion A from the literature.\label{tab:sources}}
\tablehead{
\colhead{$\alpha$} & \colhead{$\delta$} &  \colhead{Ref.\tablenotemark{a}} &  \colhead{Gaia DR2} &  \colhead{$\mu_\alpha$} & \colhead{$\mu_\delta$} &  \colhead{$\pi$} &  \colhead{RV\tablenotemark{b}}  &  \colhead{High} &  \colhead{Simbad} \\
\colhead{(J2000)} & \colhead{(J2000)} &  \colhead{} &  \colhead{Source ID} &  \colhead{(\masyr)} & \colhead{(\masyr)} &  \colhead{(mas)} &  \colhead{(\kms)}  &  \colhead{PM?} &  \colhead{ID} 
}|
\startdata
83.1837 & -5.4979 & 1 & 3209418886078218240 & $4.429 \pm 0.256$ & $-13.911 \pm 0.214$ & $2.473 \pm 0.148$ &  & r & [RHS2000] 1-121 \\
 83.7948 & -6.5790 & 4 9 & 3016942018355671168 & $-7.525 \pm 0.131$ & $-0.775 \pm 0.126$ & $3.476 \pm 0.071$ & $48.742 \pm 2.142$ & n & BD-06  1239 \\
83.3687 & -6.0351 & 1 5 & 3017208241906028032 &  1.263 $\pm$ 0.141 &  0.447 $\pm$ 0.127 & 2.167 $\pm$ 0.084 &  26.551 $\pm$ 1.581 &  & V723 Ori \\
\enddata
\tablenotetext{}{Only a portion shown here. Full table is available in an electronic form.}
\tablenotetext{a}{{1: \cite{kounkel2018a} 2: \cite{grosschedl2019} 3: \cite{fang2009} 4: \cite{fang2013} 5: \cite{hsu2012} 6: \cite{hsu2013} 7: \cite{kounkel2016} 8: \cite{hasenberger2016} 9: \cite{pillitteri2013} 10: \cite{fang2017}17} 11: \cite{megeath2012} 12: \cite{kounkel2017a} 13: \cite{rebull2006} 14: \cite{furesz2008} 15: \cite{getman2005} 16: \cite{hillenbrand1997} 17: \cite{da-rio2010} 18: \cite{da-rio2012} 19: \cite{kuhn2014} 20: \cite{getman2014} 21: \cite{getman2014a} 22: \cite{sicilia-aguilar2005}}
\tablenotetext{b}{Weighted average from \cite{kounkel2016} and \cite{kounkel2019}}
\end{deluxetable*}

\begin{figure}
\plotone{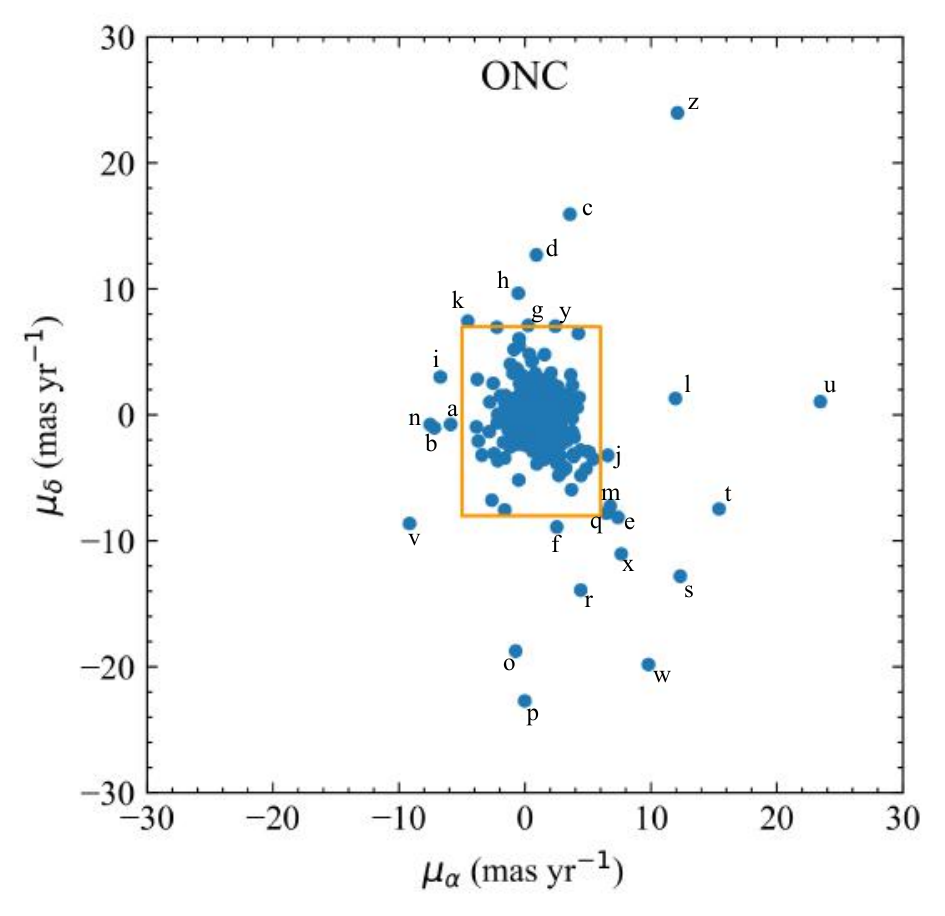}
\caption{Proper motion distributions from observations of the ONC \deleted{(left), and from a typical synthetic cluster embedded in a uniform background (right).} The box shows the cut to select high proper motion sources. The labels identify the high proper motion sources in the description throughout Section \ref{sec:pms}. At the distance of the ONC, 1 \masyr $\sim$ 2 \kms.}
\label{fig:PMs}
\end{figure}

\deleted{We projected the position of the Gaia sources backwards in time using their proper motions over a timescale of 1 Myr, with a step size of one hundred years. To test the significance of these pairs, we constructed several synthetic clusters in which the positions and proper motions of stars were drawn from a normal distribution consistent to what is observed towards the ONC in all dimensions, with a comparable number of stars, embedded in a uniform distribution of field stars [see][for a detailed description][kounkel2018a]. In all of the synthetic clusters, the frequency of these apparent 'encounters' was comparable to what was observed in the real data, nor were there any kinematical signatures that could further distinguish them. Thus, none of the identified pairs exhibit statistically significant signatures that could unequivocally identify sources that were a part of a 3-body interaction event, not if the resulting velocities  are within the velocity dispersion of the cluster. While there are likely to be a significant number of stars that have been ejected from unstable triple systems, most of them are expected to have have ejection velocities $<$2 \kms\ [reipurth2010], and thus are unidentifiable with this method. Moreover, moving through the potential well would cause these stars to accelerate, requiring a more complex model to evaluate possible pairs.
However, one significant difference between the real data and the synthetic clusters is the presence of sources with high velocity proper motions, in excess of 10 \kms\ from the cluster mean. The only such sources in the synthetic sample are the foreground and the background stars, not the real cluster members (Figure \ref{fig:PMs}).}

\added{Most stars in the ONC are dynamically cold, well within the dispersion velocity of the cluster. While there are likely to be a significant number of stars that have been ejected from unstable triple systems, most of them are expected to have have ejection velocities $<$2 \kms\ [reipurth2010], thus they are difficult to distinguish from other stars that did not have such a dynamical interaction. Moreover, simply moving through the potential well would cause these stars to accelerate, thus, any signature that the ejection might have imprinted on their kinematics would quickly become even less pronounced.

However, some sources towards the ONC have high velocity proper motions, in excess of 10 \kms\ from the cluster mean (Figure \ref{fig:PMs}). Excluding the contamination from the old field stars that do have large velocity dispersion, for the young stars to achieve such speeds, a strong dynamical scattering would be required.}

The filters described above are effective at separating out older stars. However, even though the original selection young stars comes from the studies of the Orion A, there may be contamination from other nearby young populations. When young stars are formed, they are dynamically cold, and the dispersion velocity of these populations remains on the order of just a few \kms\ for several tens if not hundreds of Myr, and the stars within them can be identified as part of a coherent comoving group \citep{kounkel2019a}. All of the young stellar populations with ages less than $\sim$70 Myr have proper motions in both $\alpha$ and $\delta$ close to zero, thus, none of them would appear to have intrinsically high proper motions in the reference frame of the ONC. Thus, for a confirmed young star to be accelerated to such speeds relative to the mean velocity, the likeliest mechanism is through a 3-body interaction that resulted in an ejection event, although some of these ejected stars may originate from the nearby populations.

We selected sources as having high proper motions if they were found outside of the box

$$-5 \masyr < \mu_\alpha < 6 \masyr$$
$$-8 \masyr < \mu_\delta < 7 \masyr$$

\begin{figure}
\begin{centering}
\includegraphics[width = \linewidth]{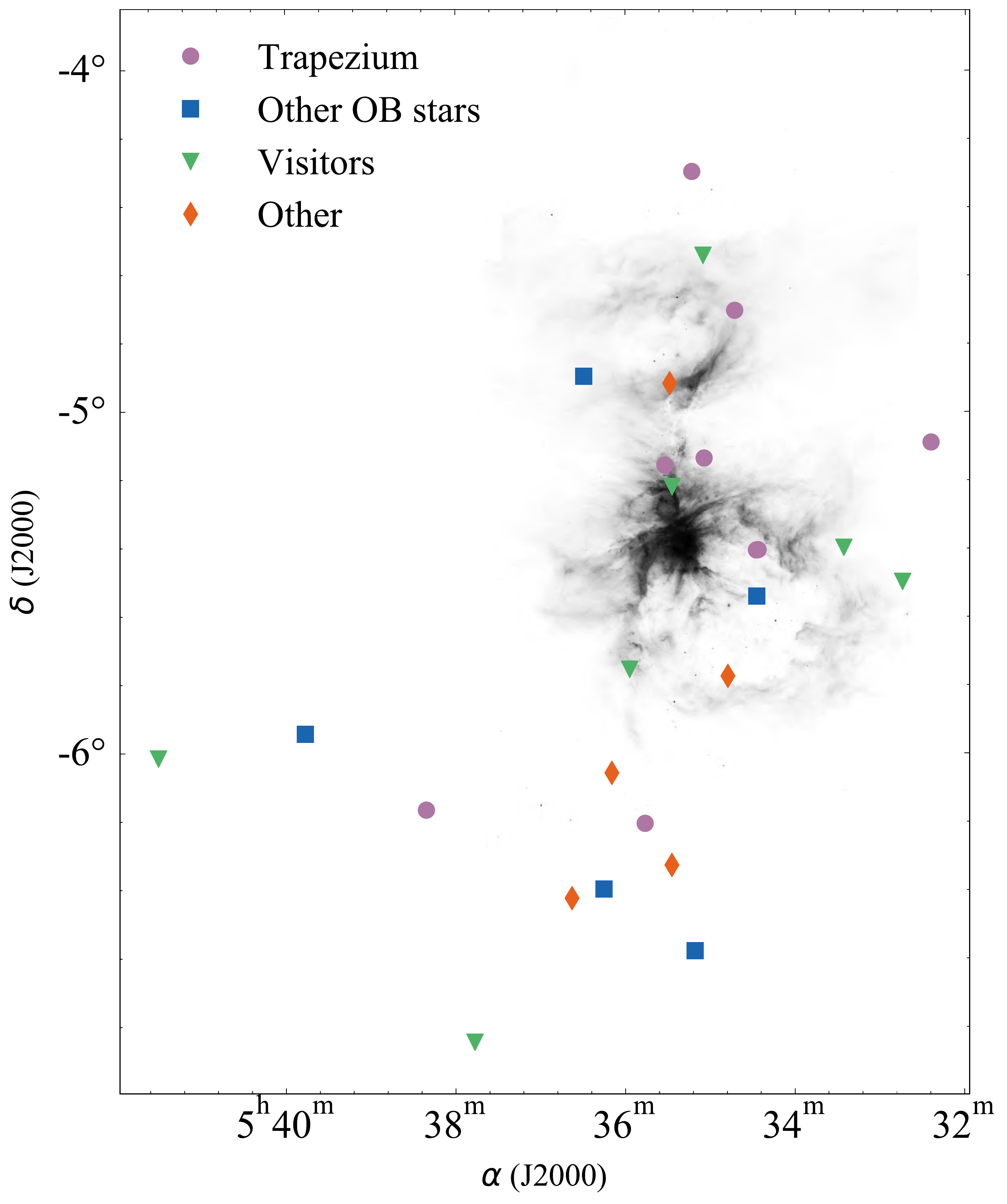}
\caption{Distribution of the high proper motion sources (colored according to their apparent point of origin), projected against the Spitzer 8$\mu$m background \citep{megeath2012}.}
\label{fig:field}
\end{centering}
\end{figure}

\noindent These cuts are able to exclude not only the sources with the kinematics within the velocity population of the ONC, but also any of the other identified populations towards the Orion Complex. This identified 26 sources (Figure \ref{fig:field}). This selection is somewhat conservative, and there are other systems outside of the dispersion velocity of the ONC that can also be considered bona fide runaways.

There are also 9 stars outside of the 2$^{\circ}$ search radius, along L1641, that meet these criteria, although we do not focus on them due to some difference in kinematics along the Orion A filament and a less comprehensive membership list. 

\section{Evaluation of candidates}\label{sec:pms}

We attempt to characterize the origin of for all of the identified high proper motion sources. Given the speed of these stars, traveling through the potential well of a cluster should not alter the measured proper motions relative to the initial ejection significantly, thus a uniform traceback is an acceptable approximation of their path. We first convert proper motions to the cluster rest frame by subtracting the average kinematics of the ONC in both $\mu_\alpha$ and $\mu_\delta$. We then project the apparent position of the stars back in time, incorporating errors in proper motions, searching for likely candidates of the interactions which have resulted in ejection. If such a candidate of origin could be identified for a given ejected star, we stop the projection at the timestep where the interaction has been likely to take place. Otherwise, the path of the star is projected back for 1 Myr. Unless otherwise specified, the spectral types are from \citet{hillenbrand1997}. We further include the discussion of the specific youth signatures of the identified sources, which include a presence of a protoplanetary disk \citep{rebull2006,megeath2012,grosschedl2019}, Li I absorption and/or accretion from H$\alpha$ emission consistent with either classical or weak-lined T Tauri star \citep[CTTS or WTTS][]{sicilia-aguilar2005,furesz2008,fang2009,fang2013,hsu2012,kounkel2016,fang2017,kounkel2017a}, X-ray properties \citep{getman2005,pillitteri2013,kuhn2014,getman2014,getman2014a,hasenberger2016}, low $\log$ g \cite{kounkel2018a} or high bollometric luminosity \citep[$L_{bol}$][]{hillenbrand1997,da-rio2010,da-rio2012,hsu2013}.

\added{Eight of the 26 identified sources have high unit weight error (RUWE) \footnote{\url{http://www.rssd.esa.int/doc_fetch.php?id=3757412}}, which shows that they are not well-fit by the astrometric five parameter solution, such as in cases if they are astrometric binaries or extended sources. These sources also have higher uncertainties in their parameters, and while in all cases they are inconsistent with having low proper motions within the cluster dispersion velocity, they should be treated with caution until \textit{Gaia} DR3. We identify them in the text for the further follow up.}

\begin{figure*}
\includegraphics[width = \linewidth]{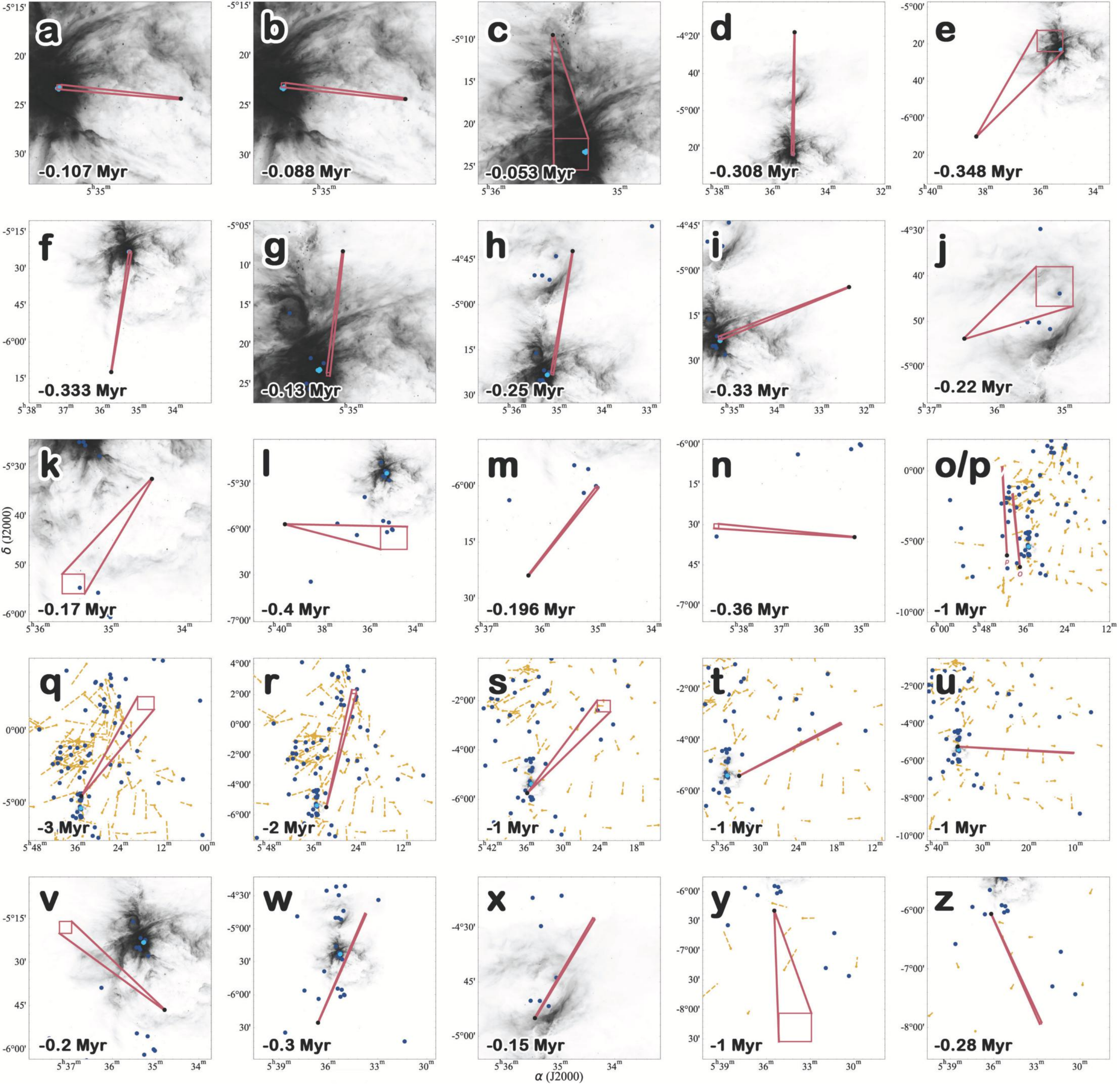}
\caption{The apparent path of the high proper motion sources projected back in time over the course of the period shown in the bottom left corner of each image. The cone shows the uncertainty in the path. The sources are projected against the Spitzer $\mu$m background \citep{megeath2012}for a reference of their position. Blue points show known OB stars in the region, cyan shows the Trapezium stars when applicable. Yellow dashed lines show the characteristic trace-back projection of the parts of the Orion Complx outside of the ONC for the corresponding duration, with the yellow dot corresponding to the current position \citep{kounkel2018a}.\label{fig:pyramid}}
\end{figure*}

\subsection{Sources originating from Trapezium}

Nine high proper motion stars appear to originate from the Trapezium region. As the central and most massive region in the ONC, it appears to be at rest in the cluster reference frame, allowing for a robust identification of it as a site of origin. This region contains several OB stars that have numerous companions \citep[e.g.][]{costero2019}, some of which could have ended up in an unstable configuration and been ejected. Moreover, as the most dynamical region in the ONC, close encounters between stars that have not been formed together as a part of a multiple system are also likely. 

Brun 259 (Figure \ref{fig:pyramid}a; K7, WTTS) and V1961 Ori (\added{Figure \ref{fig:pyramid}}b; G9IV-V, Li I absorption) are both notable, as they both have similar trajectories, comparable speeds ($\sim$10 \kms), and separated by only 17''. Most likely they have been ejected concurrently, $\sim$0.1 Myr ago. V1961 Ori has been previously identified as a runaway \citep{kounkel2017}, and is also a known spectroscopic binary \citep{kounkel2019}.

V360 Ori \citep[\added{Figure \ref{fig:pyramid}}c; known binary, spectral types M0.5+M1.5][accreting disk, \added{high RUWE}]{daemgen2012} has been ejected most recently out of the stars in the sample ($\sim$0.05 Myr ago). The remaining three -- 2MASS J05351295-0417499 (\added{Figure \ref{fig:pyramid}}d low $\log$ g), 2MASS J05382070-0610007 (\added{Figure \ref{fig:pyramid}}e, disk-bearing, \added{high RUWE}), and Haro 4-379 \citep[\added{Figure \ref{fig:pyramid}}f, K7.5][X-ray and Li I]{hsu2012} -- would have had a strong encounter over 0.3 Myr ago.

V1321 Ori \citep[\added{Figure \ref{fig:pyramid}}g, previously identified runaway,][K0V,Li I]{kounkel2017}, V1440 Ori (\added{Figure \ref{fig:pyramid}}h, Li I), and CRTS J053223.9-050523) \citep[\added{Figure \ref{fig:pyramid}}i, M2][disk-bearing]{rebull2000} do not project directly back to the $\theta^1$ Ori stars of the Trapezium in the cluster reference frame, but are likely to have originated from this region as well. 

\subsection{Interactions With other OB Stars}

When a path of an ejected star is projected back in time, an OB star along that path is more likely to be a progenitor compared to a low mass star. This is because, due to their mass, OB stars have a higher multiplicity fraction, and an interaction with a massive system is more likely to impart a higher ejection velocity to the companion. Additionally, an OB star would have a smaller kickback velocity. While they do not necessarily have to be entirely comoving with the cluster mean velocity, because they are likely to sit somewhat deeper in the potential well, we treat them as stationary. 

V836 Ori \citep[\added{Figure \ref{fig:pyramid}}j, M3.7][accreting disk, \added{high RUWE}]{fang2017} can be directly projected to the OB star HD36958 near the NGC 1977 region.

Several stars appear to originate from NGC 1980 region in the south of the ONC. V1116 Ori (\added{Figure \ref{fig:pyramid}}k, M3, low $\log$ g and high $L_{bol}$, \added{high RUWE}) can be projected to $\iota$ Ori. ESO-HA 1713 (\added{Figure \ref{fig:pyramid}}l, disk-bearing, \added{high RUWE}) could potentially have interacted with one of several OB stars in this region - HD36959, HD36960, HD37025, or HD37209. Parenago 2374 (\added{Figure \ref{fig:pyramid}}m high $L_{bol}$) projects to HD36959.

BD-06 1239 \citep[\added{Figure \ref{fig:pyramid}}n, F5.5][Li I and X-ray]{fang2013} comes from L1641 region, near the vicinity of HD 37481.

\subsection{Visitors to the ONC}

Several stars have very high proper motions, even compared to the rest of the sample, and they tend to move towards the ONC, not away from it. For us to see them in the vicinity of the ONC would imply that they either have just been ejected, or that they originate from outside of the ONC, from the nearby star forming regions. While it is difficult to determine the exact system from which they have been ejected, we can estimate the general region from which they come from.

The sources in Parenago 2600 \citep[\added{Figure \ref{fig:pyramid}}o, F8.5][high $L_{bol}$]{hsu2013} and Gaia DR2 3017044689550345856 (\added{Figure \ref{fig:pyramid}}p, low log g) both appear to be coming from either $\sigma$ Ori cluster, or from Ori OB1b region, likely to have been ejected $\sim$ 1 Myr ago.

A number of stars appear to originate from Orion D: sources 2MASS J05350504-0432334 (\added{Figure \ref{fig:pyramid}}q, low log g, WTTS, \added{high RUWE}) and 2MASS J05324407-0529523 \citep[\added{Figure \ref{fig:pyramid}}r, a known spectroscopic binary][low log g]{kounkel2019}, are likely to originate from the northern part of the Ori OB1a, possibly from the 25 Ori cluster, $\sim$1.5--2 Myr ago. While V1589 Ori (\added{Figure \ref{fig:pyramid}}s, accreting disk, \added{high RUWE}) passes near the dense core of the cluster, it is more likely to originate from $\eta$ Ori. 2MASS J05332561-0523541 (\added{Figure \ref{fig:pyramid}}t WTTS) and Brun 711 (\added{Figure \ref{fig:pyramid}}u, G3IV-V, WTTS) appear to come from the mid-to-south portion of the Orion D.

\subsection{Remaining Sources}

For the remaining 5 stars, we are unable to identify a definitive point of origin, as they could have interacted with any of the low mass stars along their projected path.

V774 Ori \citep[\added{Figure \ref{fig:pyramid}}v, M1.3e][Li I and X-ray]{kounkel2017a}, and BD-06 1256 \citep[\added{Figure \ref{fig:pyramid}}w, A8.0][Li I and X-ray]{hsu2013} probably originate from near the center of the ONC. BD-06 1256 appears to be the most massive of the high proper motion sample from its placement on the HR diagram.

V1739 Ori (\added{Figure \ref{fig:pyramid}}x low log g) could have potentially interacted with several stars in the NGC 1977 region without definitively projecting to any massive stars, or else originate from outside of the cluster. V799 Ori \citep[\added{Figure \ref{fig:pyramid}}y, M1.5][disk-bearing WTTS with X-ray, \added{high RUWE}]{fang2009} is originating from central L1641. Both are known spectroscopic binaries \citep{kounkel2019}.

2MASS J05360962-0603316 (\added{Figure \ref{fig:pyramid}}z, WTTS) is the fastest star identified, with a velocity of approximately 58 \kms. It has either just recently been ejected locally from the northern portion of L1641, or possibly from outside of the Orion Complex, although it is unclear as to from where. 

\section{Discussion and Conclusions}

Using \textit{Gaia} DR2, we have identified 26 stars in the vicinity of the ONC that were involved in a 3-body ejection event within the previous 1-2 Myr. Nine of them can be projected back towards the Trapezium, the most dynamical region of the ONC, 5 stars can be projected back towards other various OB stars in the region, and 5 stars do not have a progenitor that could be definitively identified. The 7 other stars, while still young, appear to originate from outside of the ONC, from other regions that are also part of the Orion Complex.

The average fraction of disk-bearing stars in the ONC in the full catalog of 5988 stars, with the sources with disks identified by \citet{megeath2012} and \citet{grosschedl2019} is $\sim$50\% \citep[which is consistent with e.g.,][]{mamajek2009}, though the extinction results in a bias against reliable measurements of parallaxes and proper motions with \textit{Gaia} compared to their diskless and less dusty counterparts. Thus out of 1871 stars in the final sample, only $\sim$40\% have disks. Out of the high proper motion stars, 7 out of 26 are disk-bearing:(V360 Ori, 2MASS J05382070-0610007, CRTS J053223.9-050523, V836 Ori, ESO-HA 1713, V1589 Ori, and V799 Ori), though, 7 do originate from the older populations that have been more likely to naturally \added{dissipate} their disks over time. Excluding these sources, 6 out of remaining 19 have disks, for a disk fraction of 32+/-13\%. Therefore, while ejected stars are somewhat more likely to be ripped from their protoplanetary disk in the process of a strong dynamical interaction compared to the stars in a more quiescent state, the difference is not stark.

All of the identified ejected stars have high proper motions, moving with speeds in excess of 10--60 \kms. Interactions that result in lower velocity ejections are more common. But even with the unprecedented precision of \textit{Gaia} DR2, if the resulting speed is within the velocity dispersion of the cluster, it is difficult to identify them in a statistical manner compared to chance alignments. To do so, it is necessary to involve distances and radial velocities in the analysis. However, precision in parallax does not compare to the precision of the position of stars in the plane of the sky, and while the ONC is one of the best surveyed young clusters with radial velocity information available for a large number of stars, the measurements are not available for all members.

Nonetheless, 26 stars that were involved in some of the most extreme dynamical interactions in the region are not an insignificant part of the total membership list, representing $\sim$1.4\% of the full sample that was analyzed. \added{Even though this census of the ejected stars may be incomplete due to the cuts imposed on the initially selected data,} this shows that 3-body interactions within dense clusters are not uncommon.

Of even greater interest are the stars that are 'visiting' the  ONC. The only reason why it was possible to identify them is because they are projected in the vicinity of this cluster and thus they have benefited from being analyzed and had their youth confirmed alongside the bona fide cluster members. But because the ejected stars do not necessarily have a preferred direction to scatter, and the area in the vicinity of the ONC is only a small patch in the surroundings of these nearby populations, it is quite likely many more stars have been scattered in the direction where they are not easily identifiable. This also may have implications for the ONC, in that there might have been several even higher velocity stars, but that they have already traveled outside of our search radius, thus our catalog is most likely not complete.

However, the ONC is one of the best studied nearby young clusters, and there were many surveys to identify its members through several different techniques. Such extensive membership lists may not necessarily be available in other star forming regions. Contamination from the older field stars is of significant concern when it comes to identifying high proper motion members, and thus prior membership information is necessary for a robust identification. With \textit{Gaia} DR2 recently there was significant success in identifying members of young populations through clustering analysis \citep[e.g.,][]{kounkel2018a}, but such an approach works only for finding stars with representative kinematics to their parent population; it is not suitable for searching for ejected stars. Similarly, with knowing distances to the individual stars it is now possible to find populations of young stars through their position on the HR diagram \citep{zari2018}, but such an approach is not entirely free of contamination. More effort would be needed to be able to robustly perform such an analysis in the future.

\section{Acknowledgements}
This work has made use of data from the European Space Agency (ESA) mission
{\it Gaia} (\url{https://www.cosmos.esa.int/gaia}), processed by the {\it Gaia}
Data Processing and Analysis Consortium (DPAC,
\url{https://www.cosmos.esa.int/web/gaia/dpac/consortium}). Funding for the DPAC
has been provided by national institutions, in particular the institutions
participating in the {\it Gaia} Multilateral Agreement.

\bibliographystyle{aasjournal.bst}

\end{document}